\documentclass{IEEEtran}

\usepackage{cite,graphicx,amsmath,amssymb,hyperref,url,xcolor,multirow}
\usepackage{algorithmicx}
\usepackage{algpseudocode}
\usepackage{cases}
\usepackage{ulem}
\usepackage{epstopdf}
\usepackage{listings}

\ifCLASSOPTIONcompsoc
\usepackage[tight,normalsize,sf,SF]{subfigure}
\else
\usepackage[tight,footnotesize]{subfigure}
\fi

\definecolor{awesome}{rgb}{1.0, 0.6, 0.0}

\lstdefinelanguage{JavaScript}{
	keywords={typeof, new, true, false, catch, function, return, null, catch, switch, var, if, in, while, do, else, case, break},
	keywordstyle=\color{blue}\bfseries,
	ndkeywords={class, export, boolean, throw, implements, import, this},
	ndkeywordstyle=\color{darkgray}\bfseries,
	identifierstyle=\color{black},
	sensitive=false,
	comment=[l]{//},
	morecomment=[s]{/*}{*/},
	commentstyle=\color{purple}\ttfamily,
	stringstyle=\color{red}\ttfamily,
	morestring=[b]',
	morestring=[b]"
}

\begin{document}

\title{Automatic Detection and Query of Wireless Spectrum Events from Streaming Data}

\author{
	\IEEEauthorblockN{Carolina Fortuna \IEEEauthorrefmark{1}\IEEEauthorrefmark{2}, Timotej Gale \IEEEauthorrefmark{2}, Toma\v z \v Solc  \IEEEauthorrefmark{2}\IEEEauthorrefmark{3}, Mihael Mohor\v ci\v c \IEEEauthorrefmark{2}\IEEEauthorrefmark{3}}\\
	\IEEEauthorblockA{\IEEEauthorrefmark{1}Jo\v zef Stefan Institute, Ljubljana, Slovenia\\
		\IEEEauthorrefmark{2}ComSensus Ltd., Ljubljana, Slovenia\\	
		\IEEEauthorrefmark{3}Jo\v zef Stefan International Postgraduate School, Ljubljana, Slovenia\\
		\{carolina.fortuna, tomaz.solc, miha.mohorcic\}@ijs.si, timotej.gale@comsensus.eu}}
	
\maketitle

\begin{abstract}
	Several alternatives for more efficient spectrum management have been proposed over the last decade, resulting in new techniques for automatic wideband spectrum sensing.
	However, while spectrum sensing technology is important, understanding, using and taking actions on this data for better spectrum and network resource management is at least equally important.
	In this paper, we propose a system that is able to automatically detect wireless spectrum events from streaming spectrum sensing data, and enables the consumption of the events as they are produced, as a statistical report or on a per-query basis. The proposed system is referred to as spectrum streamer and is wireless technology agnostic, scalable, able to deliver actionable information to humans and machines and also enables application development by custom querying of the detected events.
\end{abstract}

%
\IEEEpeerreviewmaketitle


\section{Introduction}
The modern information society is changing with the increased penetration of data driven knowledge technologies. For instance, in wireless networks, spectrum sensing hardware and algorithms for dynamic spectrum access have been thoroughly investigated \cite{zhao2007survey}. Devices implementing such technologies were foreseen to use the \textit{information} provided by the algorithms to guide the timing and configuration of their transmissions.  These technologies were, in the first phase, developed and tested in mostly limited, laboratory use cases. However, they enabled conducting long term spectrum usage studies around the world \cite{hoyhtya2016spectrum}. Such studies generated additional knowledge, on a larger scale than previously possible, but still not sufficient to draw strong conclusions according to \cite{hoyhtya2016spectrum}. More recently, broadband multi-GHz real-time analytics enables \textit{fast} generation of information by guiding the sensing devices \cite{shi2015beyond}. Furthermore, real-time wideband spectrum sensing systems able to monitor larger portion of the spectrum are being proposed \cite{qin2016wideband}. 

However, while spectrum sensing technology is important, understanding, using and taking actions on this data for better spectrum and network resource management is at least equally important if not more. Existing spectrum measurement systems are not optimized to detect specific signals in each band \cite{hoyhtya2016spectrum}. Therefore, approaches for easier detection and exploration of real-time data streams, automatic detection, classification, and querying the information generated are all emerging research topics. Advanced algorithms, protocols and system architectures they may eventually lead to true self-optimized and self-managed wireless networks. To this end, machine vision algorithms or, more recent, deep learning techniques appear as a natural choice to automatically detect transmission blocks. However, most of such algorithms are unable to perform the detection on streaming data, are computationally expensive or require large amounts of labelled data for training a classification model. This training data, also called the ground truth, is typically manually produced, thus slow and expensive to acquire.

In this paper, we propose a spectrum streamer, i.e. a system that is able to automatically detect wireless spectrum events from streaming data, and enables the consumption of the events as they are produced, as a statistical report or on a per-query basis. By wireless spectrum events, we refer to basic events such as \textit{transmission start} and \textit{transmission stop}. The proposed system uses time series algorithms implemented in an efficient and scalable approach in line with state of the art stream processing systems and does not require labels for training nor any apriori models. However, the evaluation of the performance of the proposed system is also depending on labelled data.

To the best of our knowledge, this is the first proposal of this kind in the literature and can be used to realize and \textit{automatize} all five spectrum measurement phases of the methodology proposed in \cite{hoyhtya2016spectrum}. The unique design principles of the proposed sprectrum streamer are:

\paragraph{Technology agnostics} being able to detect any kind of technology such as standard compliant or non-standard signals. The only assumption is that a metric reflecting spectral activity, such as detected energy, is reflected in the ingested data. 
\paragraph{Scalability} being able to balance the computation load by distributing the pipelines across various machines available in a cloud facility, be it remote or close to the edge. By using advanced principles from state of the art stream processing engines \cite{kulkarni2015twitter} able to process hundreds of millions of messages daily, the proposed system can run on a single machine monitoring band on the order of MHz but it is also able to scale to several machines and several GHz of spectrum.
\paragraph{Automatic delivery of near-real-time spectrum event information} pushing the results of the detection to interested actors such as machines and humans. Spectrum brokers, spectrum databases or machines can use the detected events to manage the wireless network. Real-time detected events can also be delivered to human users for visualization, however the amount of information can be overwhelming. 
\paragraph{Delivery of historical reports} supporting computing statistics of the detected events. The historical reports can also be used by machines or network management systems that employ a reactive management strategy, however they are more suitable to human users that are unable to process large amounts of fast information. The statistical reports can be used as an enabler for automatically collecting spectrum usage data for studies such as in \cite{yin2012mining} or as a report generator for regulators and other interested bodies.
\paragraph{Custom application development} allowing developers to write custom programs that are able to query the detected events by frequency, time and location using efficient indexes. In such way, time/space/frequency studies of the detected events are enabled.

This paper is structured as follows. Section \ref{sec:statement} states the event detection problem. Section \ref{sec:solution} presents the proposed stream detection system. Section \ref{sec:implementation} details the reference implementation of the spectrum streamer, and Section \ref{sec:eval} reports on the evaluation methodology and results. Section \ref{sec:related} identifies ans summarizes related work while Section \ref{sec:conclusions} concludes the paper and outlines future research directions.

\section{Problem statement}
\label{sec:statement}

\begin{figure*}
	\subfigure[Spectrogram with example transmissions from two different technologies in the ISM 868 MHz band.\label{fig:spectrum}]{
		\includegraphics[width=0.48\linewidth]{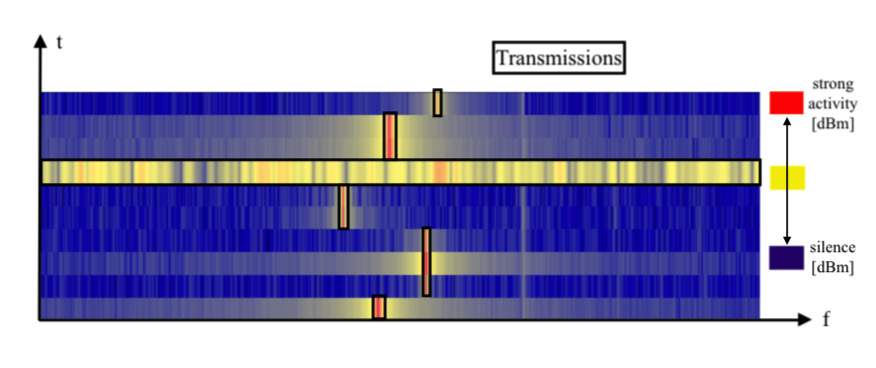}
	}
	\subfigure[Formalization of the problem.\label{fig:formal}]{
		\includegraphics[width=0.48\linewidth]{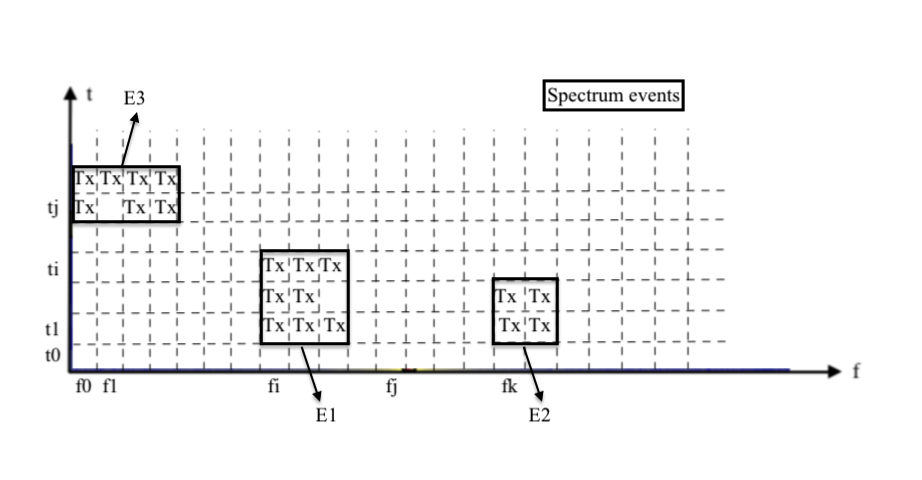}
	}
	\caption{Intuitions for the proposed event detection system.}
\end{figure*}

In this work, we aim to automatically detect events that represent transmission blocks in wireless radio spectrum, without any prior knowledge of how these transmissions might look like. Figure \ref{fig:spectrum} presents a snapshot of spectrum in the 868 MHz band. The horizontal axis represents the frequency, the vertical axis the time while the color represents the energy measured in that band. Red color means that a strong transmission exists, yellow stands for a weaker transmission, possibly from further away while blue stands for silence (i.e. absence of radio signal). There are 6 transmission blocks, enclosed in black rectangles, in Figure \ref{fig:spectrum}.  Five transmission blocks are from a narrowband Random Frequency Division Multiple Access (RFDMA) system and one comes from a IEEE 802.15.4 system. Given this example figure, we aim to automatically detect and describe the transmission events $E_i$:
\begin{equation}
\label{eq:tuple}
<E_i, t_{start}, t_{stop}, f_{start}, f_{stop}>
\end{equation}

Additionally, we aim to perform this detection on streaming data as it comes from available off-the-shelf or advanced spectrum sensing devices \cite{nafkha2014experimental,qin2016wideband} in the form of power spectral density. This means that, at time $t_i$, determined by the sampling frequency, a vector $S_i$ of energy levels $e_i$ corresponding to each frequency bin $f_i$ (i.e. the smallest resolution in frequency domain obtained by FFT) from the sensed spectrum becomes available:
\begin{equation}
\label{eq:sample}
S_i = [(f_1, e_1), (f_2, e_2), ..... (f_N, e_N)]
\end{equation}

Looking at Figure \ref{fig:formal}, at $t_0$, all the $e_i$ in Eq. \ref*{eq:sample} would be very low, corresponding to silence while at $t_1$, the energy levels corresponding to $f_1, f_{i+1}, f_{i+2}$  and $f_k, f_{k+1}$ would be high, corresponding to transmission activity. The only assumption we take here is that there is only one raw value per $f$ coming from the spectrum sensing system. The values can be energy levels, binary values or any other measure that corresponds to spectral activity.

As depicted in Figure \ref{fig:spectrum}, the values of $e_i$, corresponding to the colors in the spectrogram, can take various values and a simple thresholding approach is rather unreliable. In order to detect the spectrum events, the system has to undertake two major steps:

\paragraph{Perform a frequency domain detection} 
When a sample $S_i$ is pushed from the spectrum sensor to the event detector, the former has to determine whether there are frequencies in that vector on which a transmission is ongoing. The system should be able to (i) discriminate between increasing energy levels that precede an actual transmission and the actual transmission to accurately determine the start frequency; (ii) detect high/moderate relatively constant energy levels that are specific to transmissions; (iii) detect the stop of the transmission and discriminate against the decreasing energy levels that follow a transmission to be able to accurately determine the stop frequency; and (iv) disregard noise.

This step provides the $f_{start}$ and $f_{stop}$ in Eq. \ref*{eq:tuple}. For example, for the first detected event the tuple becomes $<E1, f_{start1}, f_{stop1}>$.

\paragraph{Perform a time domain detection}
As transmissions are detected for each $S_i, i=1, ...N$, the system has to determine at what time a certain transmission block started and ended. It can do this by aligning subsequent tuples detected at the previous step. Assuming at $t=i-1$, the previous step detected two events $<E_1, f_{start1}, f_{stop1}>$ and $<E_2, f_{start2}, f_{stop2}>$ while at time $t=i$, the system only detected 1 event $<E_3, f_{start3}, f_{stop3}>$, the system has to decide whether $E_3$ is a continuation of $E_1$ or $E_2$, or a completely independent event. If it is a continuation, than it has to be merged into an existing event. If it is a new transmission, then it probably means that $E_1$ and $E_2$ stopped at time $t=i$ so they get assigned the $t_{stop}$ value in the tuple, while $E_3$ gets assigned the $t_{start}=i$ in the tuple.

This step can provide the $t_{start}$ and $t_{stop}$ in Eq.\ref*{eq:tuple}. For example, looking at Figure \ref{fig:formal}, the result of this step should be complete tuples describing events such as $E_1$, $E_2$ and $E_3$.

\section{Proposed solution for automatic event detection}
\label{sec:solution}
We propose a time series processing inspired method for performing automatic event detection on streams of data. We define a vector $T(f_i)$ of time series, each time series corresponding to a frequency bin, i.e. smallest frequency resolution in the spectrum that is being monitored defined by the size of FFT. On each time series, we define two empirical distributions: $W_r$ that represents the distribution of energy levels $e_i$ in the current time window and $W_h$ that represents the distribution of energy levels on a historical time window and is significantly longer in duration than $W_r$. The historical distribution keeps a statistical model of the usual state of the channel, while the recent distribution keeps a model of the recent state and is aimed at detecting spectrum activity that is different from the normal. A metric comparing the two distributions then decides whether there is a transmission $T_x(t_i, f_j)$ on a particular frequency bin at a particular time.  $T_x(t_i, f_j)$ is represented as a small square in Figure \ref{fig:formal} while a group of $T_x$ represent an event $E_i$.

\begin{figure}[htb]
	\centering
	\includegraphics[width=\columnwidth]{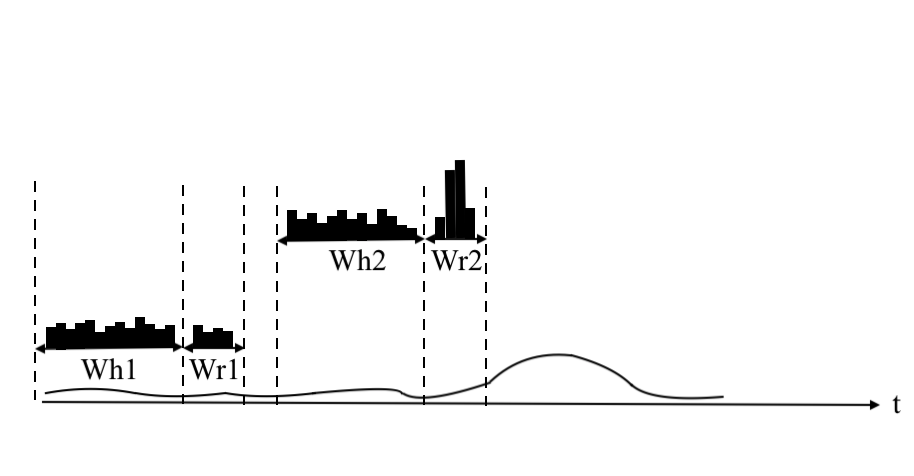}
	\caption{Statistical activity detection.}
	\label{fig:statistics}
\end{figure}

Figure \ref{fig:statistics} first depicts an example when the historical $W_{h1}$ and recent $W_{r1}$ distributions are very similar and thus no activity is detected. This corresponds to the empty squares in Figure \ref{fig:formal}. The second example in Figure \ref{fig:statistics} depicts the case where a transmission occurs on the given frequency bin. In this case, the recent distribution $W_{r2}$ is significantly different than the historical one $W_{r2}$. Their comparison, should result in activity detection and therefore a $T_x$ labelled square in Figure \ref{fig:formal}.

The main assumption of the proposed solution is that there is more silence than transmissions in the spectrum. This assumption is confirmed by several existing studies \cite{mchenry2006chicago,islam2008spectrum,yin2012mining,hoyhtya2016spectrum} that have shown that spectrum tends to be mostly unoccupied in time and/or frequency. However, our approach of detecting activity can be easily reformulated for the complementary situation where there is a lot of activity and very few silence zones. In this case, the solution would detect the silence zones that separate the transmissions and form events based on the silence zone features.

\section{Reference implementation of spectrum streamer}
\label{sec:implementation}
\begin{figure*}[htb]
	\centering
	\includegraphics[width=0.7\linewidth]{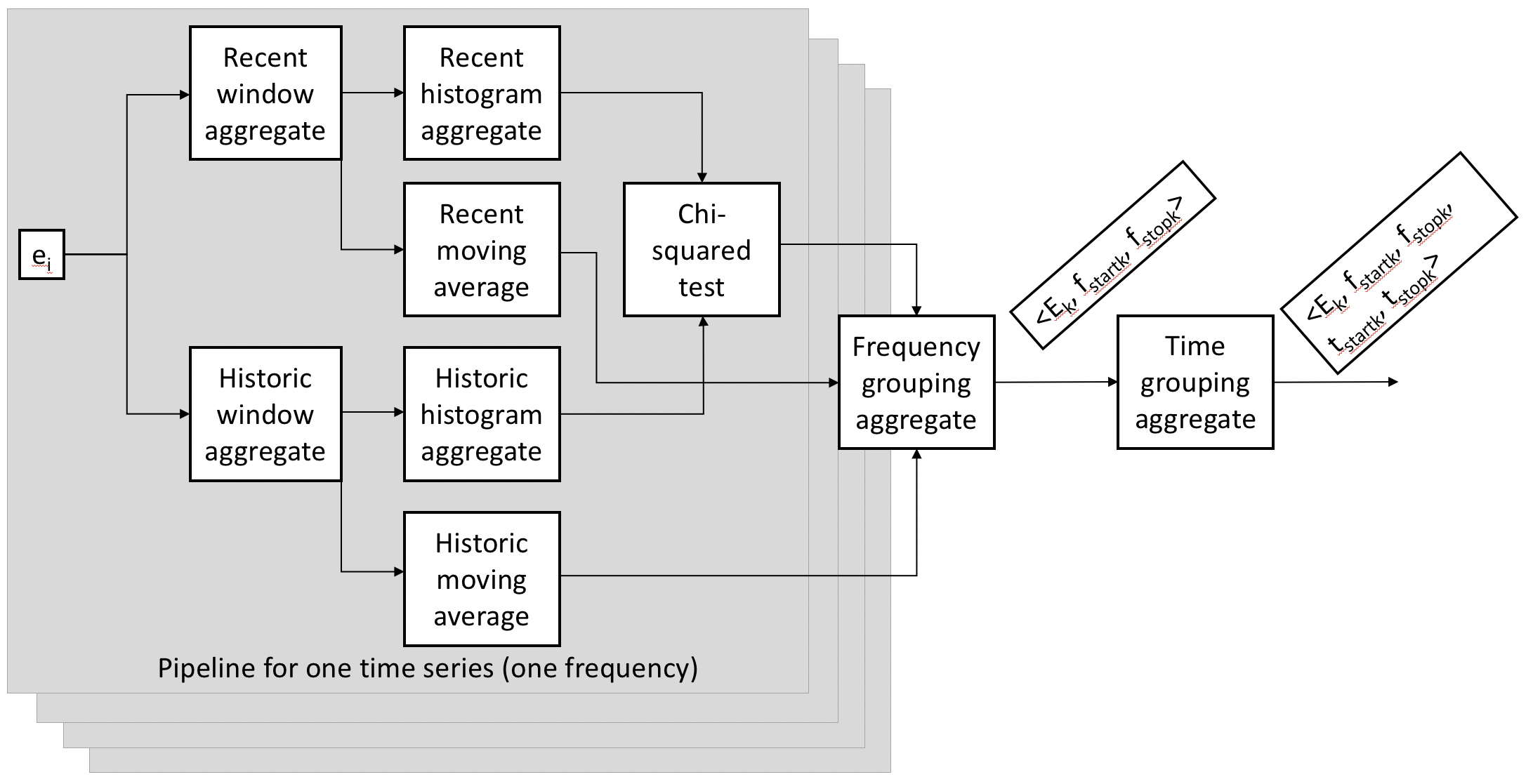}
	\caption{The event detection system architecture.}
	\label{fig:pipelines}
\end{figure*}

The reference implementation of spectrum streamer for the wireless spectrum event detection uses a stream processing approach implemented with composable pipelines. The architecture of the system is depicted in Fig. \ref{fig:pipelines}. The shaded gray area in the figure represents a time series processing pipeline corresponding to one frequency bin. The number of pipelines equals the number of elements of the sample vector $S$ from Eq. \ref{eq:sample}. On each time series, a recent and a historic windows are kept. On these windows, the corresponding histograms and moving averages are computed. The recent histogram corresponds to the distribution $W_r$  and the historic histogram correspond to the distribution $W_h$ in Section \ref{sec:solution}. The two distributions are then subject to the Chi-squared statistical test that detects whether there is any activity on that frequency bin at that timestamp. This block detects $T_x(t_i, f_j)$ from Section \ref{sec:solution} and Figure \ref{fig:formal}.

The frequency grouping aggregate then receives the results of the Chi-square tests from all the frequency bins. This module groups transmissions on adjacent frequency bins together as discussed in Section \ref{sec:statement}. The time grouping aggregate then groups transmissions in time, as they arrive, and generates the final form of the event. The time and frequency grouping aggregates are realized using simple rules.

The reference implementation of spectrum streamer uses the QMiner \footnote{\url{https://qminer.github.io/schola}} open source library that is implemented in C++ and has a thin Javascript layer that enables easy writing of applications \cite{fortuna2014qminer}. QMiner is designed to enable offline data analytics on pre-recorded batches of data or online analytics on data streams. The second approach is used in this implementation. In the QMiner architecture, the stream aggregates depicted in Figure \ref{fig:pipelines} have to be subscribed to a data store. The data store is like a table in a standard database. The role of this store in the stream processing approach of QMiner is two fold; it ensures input data compliance with a pre-defined schema and abstracts the pipeline triggering, as when a new measurements is pushed to the data store, all the subscribed pipelines are automatically triggered. In stream processing mode, the data stores do not need to store any data, thus the memory footprint is minimized to the object itself.

The pipelines approach to the event detection on streaming data is scalable to large number of frequency bins. In our implementation, since we are monitoring the frequency band between ffff and ffff, the prototype is running 1200 parallel pipelines.

\subsection{The  detection algorithm}
\label{sec:det_alg}
When implementing a pipeline that detects activity on a particular frequency, depicted in the grey area in Figure \ref{fig:pipelines}, first a data store that specifies the schema expected from the input data and triggers the pipeline was defined as presented in Listing \ref{lst:instore}. We defined a simple time series schema with $<time, value>$. 
\begin{lstlisting}[caption=Store that holds one point for each time series., label=lst:instore]
baseDef.schema.push({
	name: "Channel" + i,
	fields: [
		{ name: "Time", type: "datetime" },
		{ name: "Measurement", type: "float" }
	]
});
\end{lstlisting}

The pipeline itself is subscribed to the data store and is defined as in Listing \ref{lst:pipeline}. QMiner already provides implementations of the time series processing building blocks that are needed. These are identified by the $type$ property in lines 2, 11, 23 and 29 of Listing \ref{lst:pipeline}. The developer only needs to connect their inputs and outputs and provide parameters. For instance, the input in the recent histogram is specified in line 12 and it consists of the previously defined aggregate object in line 1. The input to the Chi-square aggregate are the recent and historic histograms as specified in lines 24 and 25. The histogram aggregates themselves are defined in lines 10 and 19 respectively. 

These parameters of each aggregate can be specified in an external configuration file and can then be tuned to improve the performance of the detection. For instance, the size of the recent histogram is set in line 5, whereas the lower and upper bounds for the dynamic histograms, the number of bins and additional bin that can be computed are set in lines 13-17.
\begin{lstlisting}[caption=The detection algorithm for each pipeline., label=lst:pipeline]
let winBuf = currentStore.addStreamAggr({
	type: 'timeSeriesWinBuf',
	timestamp: 'Time',
	value: 'Measurement',
	winsize: recentWinSize
});
let winBufDelay = currentStore.addStreamAggr({
	...
});
let hist = currentStore.addStreamAggr({
	type: 'onlineHistogram',
	inAggr: winBuf,
	lowerBound: lowerHistBound,
	upperBound: upperHistBound,
	bins: numHistBins,
	addNegInf: addNegativeBin,
	addPosInf: addNegativeBin
});
let histDelay = currentStore.addStreamAggr({
	...
});
let chi2 = currentStore.addStreamAggr({
	type: 'chiSquare',
	inAggrX: hist,
	inAggrY: histDelay,
	degreesOfFreedom: numHistBins - 1
});
let ma = currentStore.addStreamAggr({
	type: 'ma',
	inAggr: winBuf
});
let maDelay = currentStore.addStreamAggr({
	...
});
\end{lstlisting}

\subsection{The  grouping algorithms}
\label{sec:grp}
The frequency and time detection discussed in Section \ref{sec:statement} are implemented as custom aggregates and their definition can be seen in Listing \ref{lst:group} lines 1-15 and 16-32 respectively. To be triggered automatically, the aggregates have to implement the $onAdd()$ function as listed in lines 3 and 31. The current implementation of the two aggregates uses simple rules for grouping. In the case of frequency grouping, two transmissions have to be at least $F$ units apart in the frequency domain to count as independent. In the case of time grouping, two transmissions have to be at least $T$ time units apart to count as independent. $F$ and $T$ are parameters that can be tuned for obtaining the best event detection. The code that executes this logic is omitted from the paper due to size limits. The aggregates are then instantiated in the init function of the system (lines 36-43) and are wired one after the other in line 45. 
\begin{lstlisting}[caption=The frequency grouping and time grouping custom aggregates., label=lst:group]
class FqGroup {	
	constructor(settings) { ... }	
	onAdd (mergedRecord) { 
	...
	res.push({
		time: correctedTime,
		iTime: this.currentIteration,
		startFq: startFq,
		stopFq: stopFq,
		txps: [txp],
		unseen: true
		});
	}
	setParams(p){ .. }
}
class TmGroup {
	constructor(opts) { ... }

	_genEvent(msg, event) {
		return {
			description: msg,
			type: 'info',
			time: event.time,			
			channel: frequencyArray[Math.floor((event.startFq + event.stopFq - 1) / 2)],		       
			lchannel: frequencyArray[event.startFq],
			rchannel: frequencyArray[event.stopFq - 1]
		};		    
	}
	_storeTransmission(event) { ... }

	onAdd()  { 
	...
	_genEvent(m, e);
	}
}

init() {
	...
	let fqBinGroup = new FqGroup({ store: base.store("Transmissions"), 
		pValIndices: pValIndices,
		txStores: txStores
	});
	let fqBinGroupAggr = new qm.StreamAggr(base, fqBinGroup, base.store("MergedTx"));

	let tmGroup = new TmGroup({fqBinGroup: fqBinGroup, txStore: base.store("Transmissions")});
	let tmGroupAggr = new qm.StreamAggr(base, tmGroup, base.store("MergedTx"));
	fqBinGroupAggr.setParams({outAggr: tmGroupAggr});   
	...
}
\end{lstlisting}

\subsection{Scalability}
To achieve scalability, the proposed system uses advanced principles from state of the art stream processing engines \cite{kulkarni2015twitter}. The particular implementation of these principles is the QTopology\footnote{https://github.com/qminer/qtopology} stream processing layer that is compatible with QMiner. This particular stream processing layer wraps individual aggregates or pipelines such as the ones presented in Fig. \ref{fig:pipelines}. It can connect all the aggregates into a so-called $topology$. For instance, $e_i$ in Fig. \ref{fig:pipelines} becomes a so-called $spout$ that reads data from external sources, such as spectrum sensing devices and emits the data into the topology. Then, the per channel pipelines and the frequency and time grouping aggregates become so-called $bolts$ that process data and further emit it into the topology. Note that all processing in topologies is done in bolts.

To distribute the system depicted in Fig. \ref{fig:pipelines} across several machines, a worker, that runs on a single server, and includes $N$ processing pipelines for a set of specified frequency units  can be instantiated. With this approach $No-of-frequency-units/N$ workers on the same number of machines are used for activity detection. Then, the frequency and time domain grouping are instantiated as another worker on another machine, if feasible or they can also be distributes across a number of machines. With this approach, automatic processing of the entire sub 6 GHz of spectrum is enabled from multiple spectrum sensing devices in parallel.

\subsection{Real-time notifications}
Real-time notifications tend to be particularly useful for agile or dynamic spectrum and network management scenarios where machines and software entities exchange information on the current state of the spectrum. The detected transmissions can be used to update spectrum occupancy databases or to directly notify other devices that a transmission is happening in a certain channel. Real-time notifications can generate a relatively large amount of data and thus overwhelm a human user.

\begin{figure}[htb]
	\centering
	\includegraphics[width=\columnwidth]{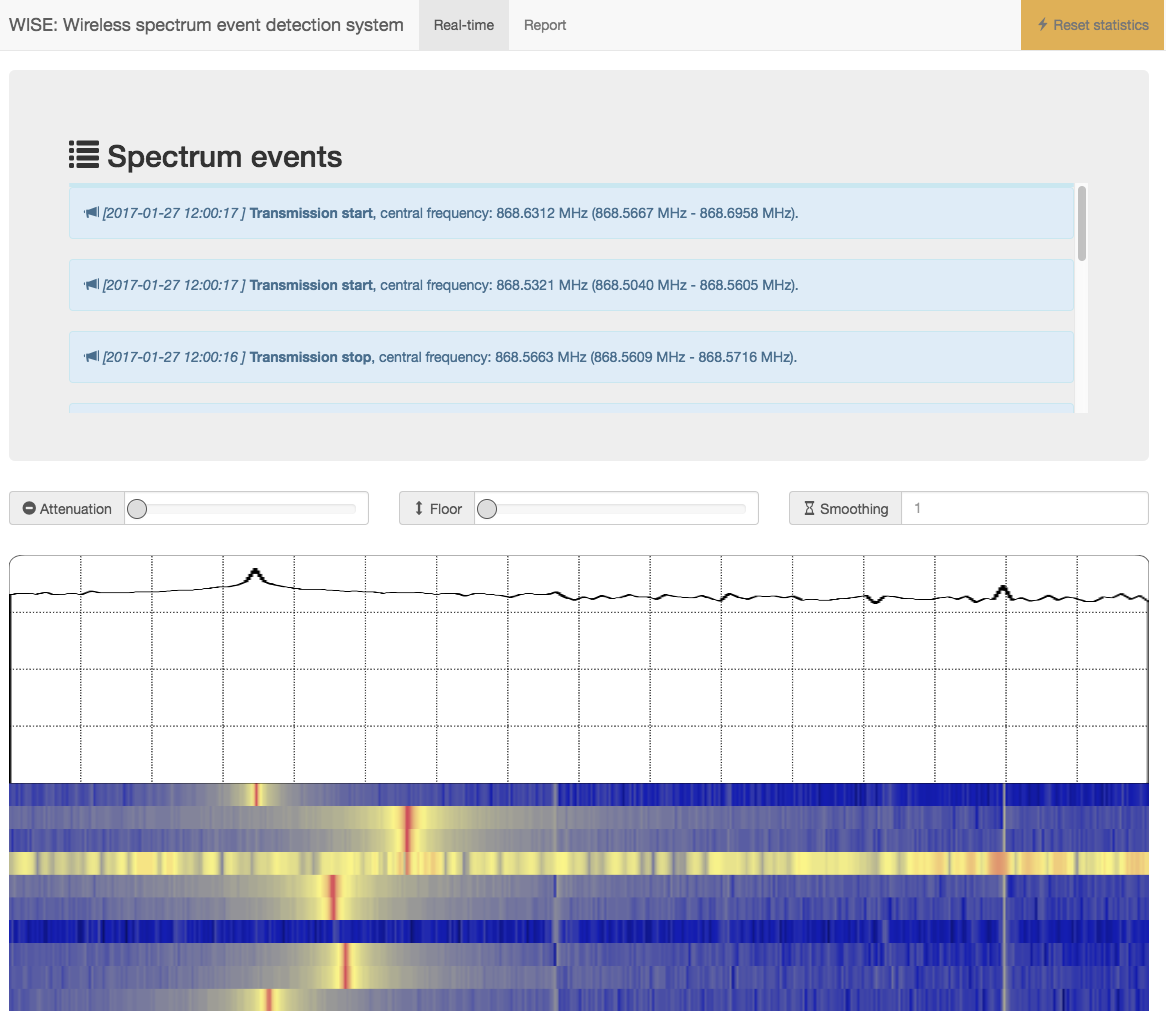}
	\caption{The prototype user interface displaying real-time events.}
	\label{fig:rt}
\end{figure}

In the reference implementation, real-time notifications are realized by sending messages to the web based user interface via WebSockets. The notifications can also be realized with other available messaging protocols such as MQTT and XMPP. For the proof of concept, with the purpose of visualizing the events and demonstrating the system, we created a user interface that displays the events. The interface is aimed as a visualization aid to help people understand what is happening in the background. The user interface is realized using web standards and widely used frameworks such as HTML5, Bootstrap and Express and the real-time notification interface is depicted in Fig. \ref{fig:rt}. 

The upper half of the user interface in Fig. \ref{fig:rt} lists the detected events while the lower part depicts the current sample and the spectrogram for a number of recent samples. It can be seen that the recent samples include two types of different transmissions.

\subsection{Statistical reporting}
Statistical reports can be delivered by computing statistics of the detected events. The historical reports can be used by machines or network management systems, but are primarily targetted to human users, for instance as a report generator for regulators or other interested bodies. 
\begin{figure}[htb]
	\centering
	\includegraphics[width=\columnwidth]{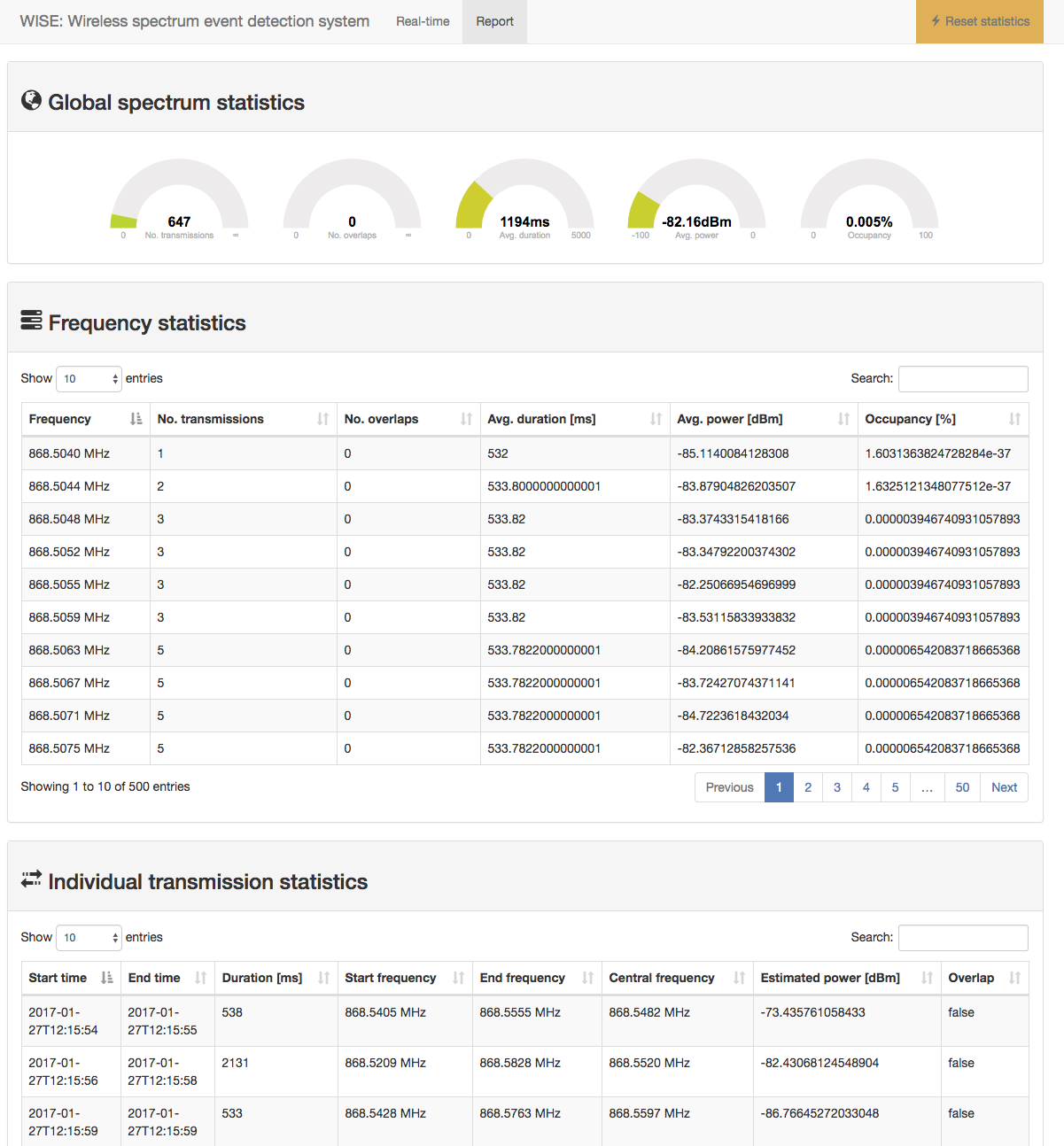}
	\caption{The prototype user interface displaying statistical reports of the events.}
	\label{fig:stats}
\end{figure}

The reference implementation of the statistical reporting uses an additional custom aggregate that counts and averages various metrics that are then displayed in the report. The realization of this report is similar to the real-time notification report and a screenshot is presented in Fig. \ref{fig:stats}. The top of the report presents global statistics over the entire monitored spectrum such as the number of transmissions, average duration of a transmission,  average power and overall spectrum occupancy. The report then continues to provide detected spectrum events per frequency unit and per channel statistics respectively. 

\subsection{Custom applications}
In addition to real-time notifications and statistical reporting, custom applications exploiting the automatically generated data can be developed and subscribed to the output of the spectrum streamer. The frequency grouped transmissions are available for querying in the $Transmissions$ data store (lines 5-12 and 39 in Listing \ref{lst:group}) while the time grouped transmissions are available in the $MergedTx$ data store (lines 21-27 and 43 in Listing \ref{lst:group}). The $time$ and $channel$ fields of the data stores have key-indexes so custom applications are able to request time-frequency related information such as how many transmissions took place at a certain hour on certain frequencies. When spectrum sensors from various locations are connected to the system, a $location$ field can be added to the two data stores with a geolocation key-index attached to it for enabling also spatial queries. This way, several types of custom applications can be built from simple statistical to more complex interference maps and hidden node detection.

Range queries \footnote{https://github.com/qminer/qminer/wiki/Query-Language} such as $<$, $\neq$ and $>$ that are most useful for the time and frequency queries can be written in a JSON-like query language as shown in Listing \ref{lst:range} or as Javascript code using filters. While range queries can also be run on location field, there exist location queries that are limited by radius (in meters) or by number of records (i.e. spectrum events) such as presented in Listing \ref{lst:location}.

\begin{lstlisting}[caption=Range queries for time\, frequency and location., label=lst:range]
{ 
	$from: <store>,
	<field1>: {$ne: <value1>}, //!=
	<field2>: {$gt: <value2>}, //>
	<field3>: {$lt: <value3>}  //<
}
\end{lstlisting}

\begin{lstlisting}[caption=Location queries., label=lst:location]
{ 
	$from: <store>,
	<field1>: { $location: [<latitude>, <longitude>], $radius: <value_in_meters> },
	<field2>: { $location: [<latitude>, <longitude>], $limit: <value> }  
}
\end{lstlisting}

\section{Evaluation}
\label{sec:eval}

For evaluating data driven systems such as the proposed spectrum streamer, we first illustrate a suitable methodology and discuss about the specifics of data preparation for evaluation and then we present the evaluation results. A transparent and clear approach to the way such systems can be evaluated is paramount for and objective and comparable evaluation of such systems.
\subsection{Evaluation methodology}
In order to evaluate the event detection spectrum streamer we employ a four step methodology as described below.

\subsection*{Step 1: Data acquisition} 
\begin{figure}[htb]
	\centering
	\includegraphics[width=\columnwidth]{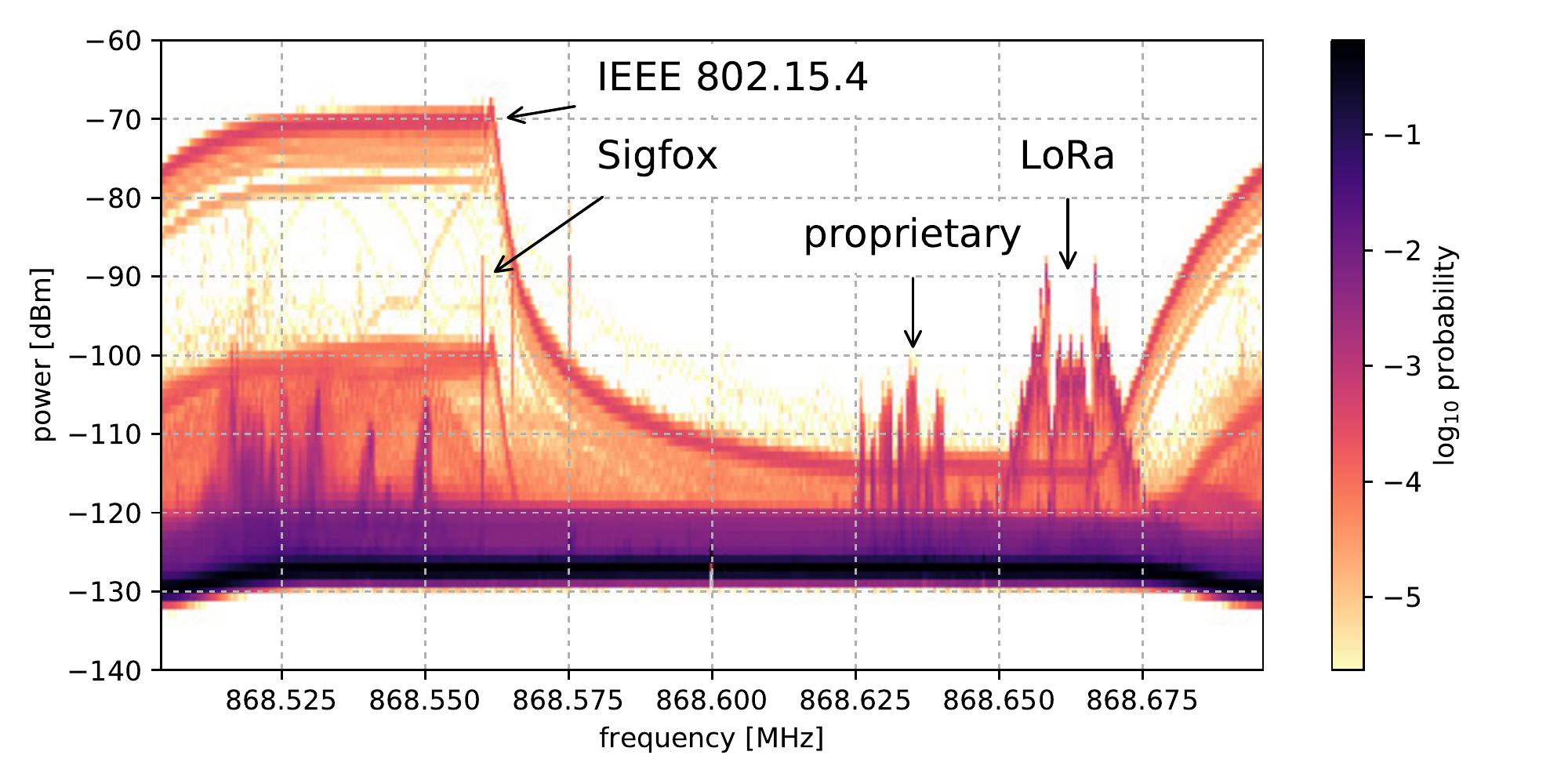}
	\caption{A histogram of power spectral density samples over a 200 kHz wide band in the unlicensed European 868 MHz SRD band in Ljubljana, Slovenia.}
	\label{fig:data}
\end{figure}

The first step in validating and evaluating the spectrum streamer is to collect data that contains spectral activity. We collected a total of 175 days (4200 hours, 75 GB in binary format) of continuous spectrum measurement data in the sub-1 GHz band. Fig. \ref{fig:data} summarizes data from the shared 200 kHz unlicensed sub-1 GHz band over a course of 24 hours showing signals from IEEE 802.15.4-based, LoRa and Sigfox networks as well as a number of unidentifiable proprietary technologies.

\subsection*{Step 2: Ground truth creation} 
The second step of the evaluation consists of creating the ground truth (or golden standard) to compare the performance of the system against. We create the ground truth by manually labelling transmission events in randomly extracted excerpts of spectrograms. Manually labelling 20 hours worth of spectrograms is difficult and time consuming. For this, we implemented a spectrum labelling script that extracts random slices of the spectrum of duration $d=10s$ that are on average $st=120s$ seconds apart. The two parameters are configurable. The manually labelled data is saved for later use.

\begin{figure*}[htb]
	\centering
	\includegraphics[width=\textwidth]{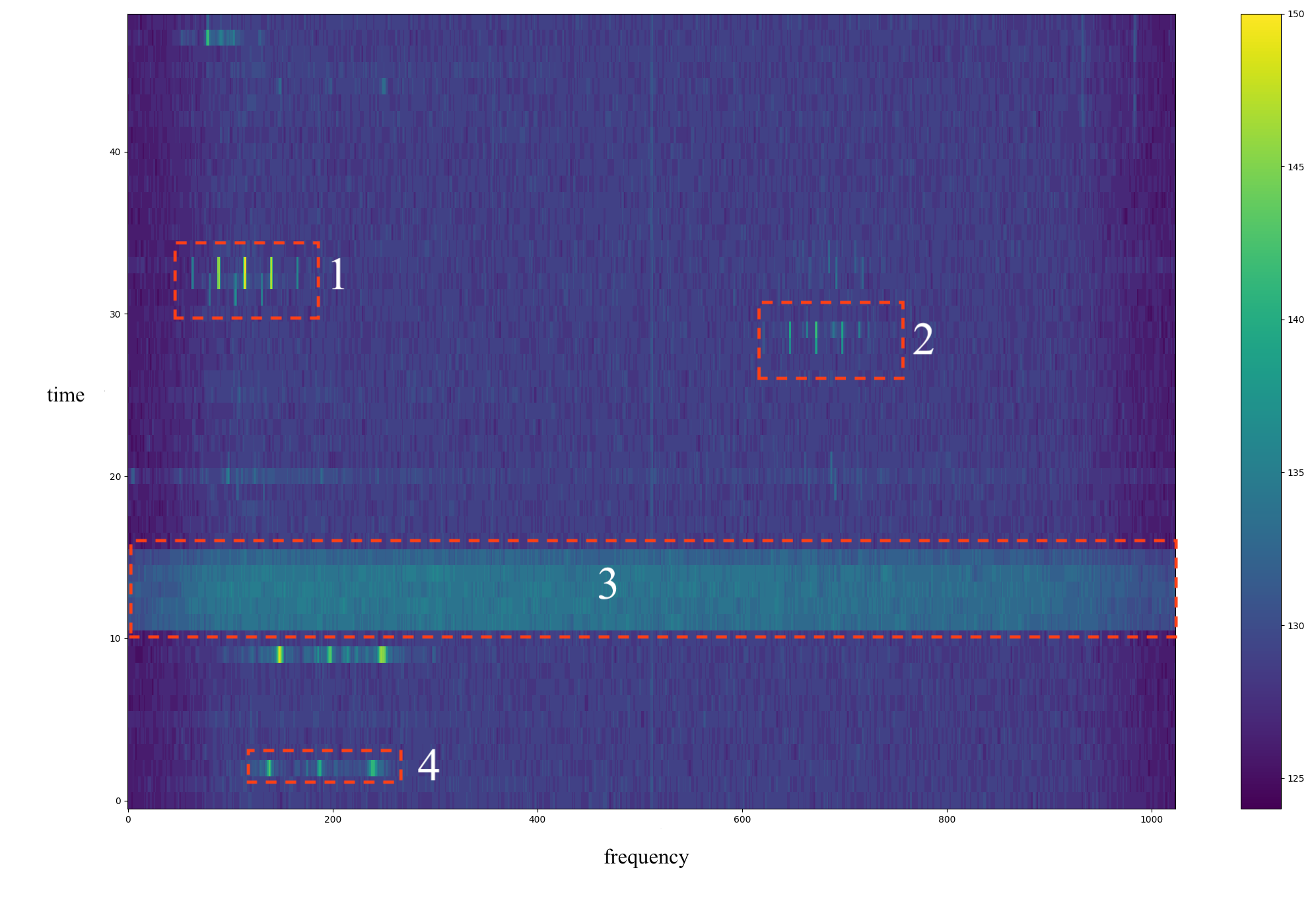}
	\caption{Random spectrogram for manual labelling.}
	\label{fig:lbl}
\end{figure*}

When creating the ground truth, the human labeler should have some knowledge about the spectral characteristics of different wireless transmissions and has to be aware of some trade-offs. For instance, presented with the spectrogram from Fig. \ref{fig:lbl}, the human labeler has to decide whether the activity in the red box 1 is a single transmission or there are 8 different transmissions. Similar decisions have to be taken for the red boxes labelled 2 and 4 while for the red box labelled 3 it is clear there is only one transmission. The labelling strategy and consistency is critical for the correct evaluation of the system. If for instance, the strategy for the activities in red boxes 1, 2 and 4 is to label 8, 7 and 3 transmissions, then the frequency grouping algorithm from Section \ref{sec:implementation} should be configured to be conservative in grouping together activities detected 2 to 3 frequency bins appart. However, if the strategy for the activities in red boxes 1, 2 and 4 is to label only 1 transmission, then the frequency grouping should be configured to be more liberal in grouping together activities detected 2 to 3, perhaps even 6 frequency bins appart.

In our evaluation, we labelled the activities in red boxes 1, 2 and 4 as a single transmission. 

\subsection*{Step 3: Automatic event detection} 
In the third step of the evaluation methodology, the automatically detected events are generated. In this step, the event detection system is triggered to run several times on the target data. Each time, it uses a different set of parameters for the sensitivity of the detection, time and frequency grouping.  The automatically detected events, for each combination of parameters, are saved for later use.

\subsection*{Step 4: Performance evaluation} 
In the fourth and last step, the automatically detected data from Step 3 is compared to the the manually labelled data from Step 2, and relevant metrics, such as the confusion matrix, are computed. The process of comparing the manually labelled data and the automatically detected events can also be prone to certain bias. Most likely, the manually labelled data and the automatic one will identify different start frequency, stop frequency, start time and stop time for the same event. So, when comparing, one has to set some thresholds, thus introducing certain bias.

For the evaluation of the system, we present the results for two scenarios as follows.

\begin{itemize}
	\item 
	\textit{Scenario 1} assumes high spectral activity for IoT ultranarrowband transmission in the considered band. This is depicted in Figure \ref{fig:spectrum} and is specific to future dense IoT networks. For this, we manually labeled 1 day of data from which we extracted 160 random 7.5 sec long snippets, totaling 20 minutes of spectrograms. The manual labelling process identified 397 events in those 20 minutes.
	\item 
	\textit{Scenario 2} assumes regular spectrum activity in a small sized European capital as depicted in Figure \ref{fig:lbl}.  For this, we are manually labelling 1 week of data from which we are extracting random 10 sec long snippets that are on average 20 minutes apart. We expect to identify about 5000 events in that week.

\end{itemize}

\subsection{Detection results}
In the following we present automatic event detection results for the two scenarios specified above.

\subsection*{Scenario 1}
We first evaluate the performance of the activity detection algorithm described in Section \ref{sec:det_alg} versus the activity detected manually. In terms of activity detection for this scenario, the proposed spectrum streamer correctly detects 87\% of the events, misses 13\% and it falsely detects 17\% as summarized in Table \ref{tab:sc1_activity}. Overall, it detects 4\% more activities than there are in the spectrum, and it incorrectly reports some events. However, this evaluation used best effort, manually selected configuration parameters. A tuning and optimization effort would significantly improve the numbers. Most of the confusions arise from the manual labelling strategy and parameter tuning discrepancies discussed in Steps 2 and 3 of the evaluation methodology. 

\begin{table}[htb]
	\renewcommand{\arraystretch}{1.1}
	\caption{Confusion matrix for the detected activity in Scenario 1.}
	\label{tab:sc1_activity}
	\centering
	\begin{tabular}{lll}
		\bfseries Correctly detected & \bfseries Undetected & \bfseries Falsely detected  \\
		[1ex] \hline \\ [-1ex]
		1517 (87\%)	&  225 (13\%)	     & 296 (17\%)	 \\
	\end{tabular}
\end{table}

Out of the 1517 correctly detected activities, it performs frequency and time grouping as described in Section \ref{sec:grp} and automatically assigns $Tx_{start}$ and $Tx_{stop}$ labels to transmission events. It correctly labels about 60\% of these events as summarized in Table \ref{tab:sc1_lbl}. The confusion comes from transmissions such as labelled with 4 in Figure \ref{fig:lbl} that is sensitive to both manual labelling and detector configuration.

\begin{table}[htb]
	\renewcommand{\arraystretch}{1.1}
	\caption{Confusion matrix for the event label assignment in Scenario 1.}
	\label{tab:sc1_lbl}
	\centering
	\begin{tabular}{l|ll}
		\bfseries  & \bfseries Tx start (manual label) & \bfseries Tx stop (manual label) \\
		[1ex] \hline \\ [-1ex]
		\textbf{Tx start (automatic)}	&  451	     & 299	 \\
		[1ex] \hline \\ [-1ex]
		\textbf{Tx stop (automatic)}	   &  329		& 438	\\	
	\end{tabular}
\end{table}

Out of the 296 falsely detected activities, 139 were labelled as $Tx_{start}$ and 157 as $Tx_{stop}$.

\subsection*{Scenario 2}

We are still computing the evaluation results for this scenario. We'll include results for 1 week, and 3 different configurations of the system.

\subsection{Computation resource considerations}
In our evaluation, we found that, on average, 1 day (24 hours) of spectrum data (200 kHz wide band in the unlicensed European 868 MHz SRD) takes 2:35 hours to process with the spectrum streamer. While processing data, the current prototype of spectrum streamer occupies less than 1 GB of memory on a MacBook with 16 GB of RAM. 

\section{Challenges}
We identify the following challenges for this line of research.

The first and major challenge is acquiring sufficient labelled spectrum data. This data would not only allow to evaluate systems such as the proposed spectrum streamer, but it would also enable labelling and training machine learning wireless technology classifiers that are becoming popular in the literature. For the time being, we are not aware of the existence of such ground truth. Obtaining manually labelled data would require a competition or a challenge where wireless engineers would use a labelling system such as the one developed in this work for evaluation. For instance, a one day hackathon at a flagship wireless networking conference might have the potential to successfully collect large labelled samples. Once a large sample of hand labelled data is available, automatic methods for labelling are being developed that are able to extrapolate it. These are the so-called weak supervision methods. 

The second challenge is to find the optimal algorithms, tradeoffs and configuration of such automatic spectrum event detection spectrum streamer and tune them to achieve high detection accuracies (above 95\%). This challenge translates into developing and evaluating algorithms for detection, and frequency, time and space grouping. The end result would be an automatic 3D spectrum event detection.

The third challenge is to engineer the final 3D spectrum detection system to perform real-time detection and scale it to cover buildings and relevant outdoor spaces. While existing stream processing and big data technologies and frameworks (Apache Spark, Twitter Heron, Cassandra, Ms StreamInsight, QMiner, etc.) are available and can be used to prototype such a system, these will not suffice. They will have to be adapted and customized for this specific application area to deal with the user and technical requirements that are relevant to the wireless community.

\section{Related work}
\label{sec:related}
Using stream analytics for developing an accurate and near real-time spectrum monitoring tool is an emerging research topic. In \cite{shi2015beyond}, the authors developed SpecInsight, a system that can very accurately schedule when, where and on which frequency band to sense to detect spectrum signals. Their system is guided by the statistical expectation that activity will be present in certain bands. Their paper draws very interesting conclusions based on extensive experimental evaluation in seven locations and a week's worth of continuous sensing at each. This work is the closest to the proposed spectrum streamer in that it focuses on streaming data. However, they use streaming data to automate the spectrum sensing process while we use streaming data to automate the sensing, detection and notifying/reporting. The spectrum streamer could for instance use the output of SpecInsight to perform the detection on large portions of the electromagnetic spectrum. 

SpecInsight is a component of the Microsoft Spectrum Observatory\footnote{\url{http://observatory.microsoftspectrum.com/}}. Other spectrum monitoring projects are the Google spectrum \footnote{\url{https://www.google.com/get/spectrumdatabase}} for measurements on TV white-spaces; the IBM Horizon \footnote{\url{https://bluehorizon.network/documentation/sdr-radio-spectrum-analysis}} project, and, more recently, Electrosense \footnote{\url{https://electrosense.org/}}. The Electrosense architecture \cite{rajendran2018electrosense} is designed for processing large volumes of batch and streaming data using popular crowdsourcing projects such as Kafka for Apache Spark for realizing real-time data pipelines, Apache Spark for real-time and batch data processing and Casandra for storage. For the time being, Electronsense does not have specialized modules implemented on top of Apache Spark to perform event detection. Such modules, if developed in the future, could be compared in detection performance and scalability to the spectrum streamer, especially considering the fact that the architecture of QMiner's stream processing functionality, on which the proposed spectrum streamer is based, is closer to Twitter Heron \cite{kulkarni2015twitter} than to Apache Spark.  

A very comprehensive and critical overview on spectrum occupancy sensing is provided in \cite{hoyhtya2016spectrum}. They show that existing spectrum usage studies generated additional knowledge, on a larger scale than previously possible, however still not sufficient to draw strong conclusions on the topic. They also propose a methodology on how to perform spectrum occupancy analysis for improving spectrum management. The system proposed in this paper can be used to realize all five phases of the methodology. 

A large body of work, as surveyed in \cite{hoyhtya2016spectrum}, focuses on centralized or distributed \cite{gorcin2014signal,pfammatter2015software} spectrum sensing techniques able to provide spectrum occupancy metrics to systems such as the one proposed in this paper.

Another related body of work relies on machine learning to identify spectral activity. The work in \cite{iyer2015detecting} uses unsupervised learning to detect and avoid interference, while  \cite{selim2017spectrum} monitors radar bands and performs classification using deep learning methods.

\section{Conclusions} 
\label{sec:conclusions}
In this paper, we proposed a system that automatically detects wireless spectrum events from streaming data and enables the consumption of the events as they are produced, as a statistical report or on a per-query basis. We formalized the event detection problem in the context of wireless communications, proposed a 3 stage solution that first detects activity and then groups this activity into frequency and time to create a full event (i.e. wireless transmission). Then we provided a reference implementation of the spectrum streamer using state of the art stream data processing tools and evaluated the system using 24 hours of continuous spectrum sensing data from the sub-1GHz spectrum band. We demonstrated that the proposed system is wireless technology agnostic, scalable, able to deliver actionable information to humans and machines, and also enables application development by custom querying of the detected events \footnote{When the paper is accepted, all the source code and data used for this paper are going to be released, so the community can repeat and use for furthering the state of the art, at: \url{https://github.com/ComSensus/spectrum_analytics}}.

While several systems for automatically collecting spectrum data are available, as we showed in the related work section of this paper, solutions for performing analytics on the data itself are rare. This work brings such a solution to the community and also identifies challenges specific to this emerging field of research.

\section*{Acknowledgment}
This work was partly funded by the Slovenian Research Agency (Grant no. P2-0016) and the European Community under the H2020 WiSHFUL (Grant no. 645 274) and NRG5 (Grant no. 762 013) projects.

\bibliographystyle{IEEEtran}
\bibliography{IEEEabrv,information}

\section*{Biographies}

\noindent
CAROLINA FORTUNA received her B.Sc. in 2006, Ph.D. in 2013 and was a postdoctoral research associate at IBCN, Ghent University, 2014-2015 and visited Stanford University for 5 months in 2017. Currently, she is research fellow at the Department of Communication Systems, Jo\v zef Stefan Institute. Her research is interdisciplinary, focusing on data and knowledge driven modelling of communication and sensor systems. She has participated in EU and industrial projects with Bloomberg LP and Siemens PSE.

\bigskip\noindent
TIMOTEJ GALE received a B.Sc. degree in Computer and Information Science from the University of Ljubljana in 2017. He is currently enrolled as a M.Sc. student at the same university and works as a student developer in ComSensus. His research is focused on data-driven technologies for the Internet of Things.

\bigskip\noindent
TOMA\v Z \v SOLC studied electronics at the Faculty of Electrical Engineering, University of Ljubljana and received his B.Sc. in 2007. He is currently a senior research assistant at the Department of Communication Systems, Jo\v zef Stefan Institute and a Ph.D. student at the Jo\v zef Stefan International Postgraduate School. His work involves hardware design, measurements and embedded software development as well as research in the field of spectrum sensing. He participated in various national and EU projects.

\bigskip\noindent
MIHAEL MOHOR\v CI\v C is head of the Department of Communication Systems and Scientific Counselor at the Jo\v zef Stefan Institute. He is also associate professor at the Jo\v zef Stefan International Postgraduate School. His research experience include development and performance evaluation of network protocols for mobile and wireless communication systems, cognitive radio networks, wireless sensor networks, dynamic composition of communication services and wireless experimental testbeds. He has participated in many EU projects. He is a Senior Member of IEEE.

\end{document}